\documentclass[twocolumn,tighten]{aastex631}
\usepackage{hyperref}
\usepackage{amsmath}
\usepackage{amssymb}
\usepackage{graphicx}
\usepackage{color}
\usepackage{subfigure}
\usepackage{bm}
\usepackage{textcomp,gensymb}
\usepackage{bigints}
\usepackage{aas_macros}
\usepackage{physics}
\usepackage{chemformula}
\usepackage{longtable}

\hypersetup{linkcolor=blue,citecolor=blue,filecolor=blue,urlcolor=blue}

\submitjournal{ApJ}

\begin{document}

\title{Revealing Limitation in the Standard Cosmological Model: A Redshift-Dependent Hubble Constant from Fast Radio Bursts}
\shorttitle{Limitation in $\Lambda$CDM cosmological model}
\shortauthors{Kalita et al.}

\correspondingauthor{Surajit Kalita}

\author[0000-0002-3818-6037]{Surajit Kalita}
\affiliation{Astronomical Observatory, University of Warsaw, Al. Ujazdowskie 4, Warsaw 00478, Poland}
\email{skalita@astrouw.edu.pl}

\author[0000-0001-8213-646X]{Akhil Uniyal}
\affiliation{Tsung-Dao Lee Institute, Shanghai Jiao Tong University, 1 Lisuo Road, Shanghai 201210, People’s Republic of China}
\email{akhil\_uniyal@sjtu.edu.cn}

\author[0000-0003-2045-4803]{Tomasz Bulik}
\affiliation{Astronomical Observatory, University of Warsaw, Al. Ujazdowskie 4, Warsaw 00478, Poland}
\email{tb@astrouw.edu.pl}

\author[0000-0002-8131-6730]{Yosuke Mizuno}
\affiliation{Tsung-Dao Lee Institute, Shanghai Jiao Tong University, 1 Lisuo Road, Shanghai 201210, People’s Republic of China}
\affiliation{School of Physics and Astronomy, Shanghai Jiao Tong University, 800 Dongchuan Road, Shanghai 200240, People’s Republic of China}
\affiliation{Key Laboratory for Particle Astrophysics and Cosmology (MOE) and Shanghai Key Laboratory for Particle Physics and Cosmology, Shanghai Jiao Tong University, 800 Dongchuan Road, Shanghai 200240, People's Republic of China}
\email{mizuno@sjtu.edu.cn}



\begin{abstract}
A major issue in contemporary cosmology is the persistent discrepancy, known as the Hubble tension, between the Hubble constant ($H_0$) estimates from local measurements and those inferred from early-Universe observations under the standard $\Lambda$ cold dark matter ($\Lambda$CDM) paradigm. Recent advances have identified fast radio bursts (FRBs), a class of extragalactic phenomena observable at considerable redshifts, as a promising observational tool for probing late-time cosmology. In this study, we incorporate two complementary methodologies, machine learning algorithms and Bayesian analysis, on a set of localized FRBs to rigorously test the consistency of the $\Lambda$CDM model at late cosmic epochs. Our results reveal a statistically significant redshift-dependent variation of $H_0$ when using separate priors on baryon density parameters $\Omega_\mathrm{b}$ or $\Omega_\mathrm{b}h^2$, indicating contradiction to the core postulate of $\Lambda$CDM. However, when the priors are combined, this redshift dependence disappears, yielding a consistent estimate of $H_0$. We further validate that the redshift dependency of $H_0$ can be removed within the more flexible framework of $w_0w_a$CDM model even without combining the priors. These findings highlight that the redshift evolution of $H_0$ is not merely an artifact of the standard model but an indication of a deeper inadequacy in the $\Lambda$CDM model, supporting the need for a more flexible cosmological framework.
\end{abstract}



\section{Introduction}

Since their initial discovery, fast radio bursts (FRBs), transient radio pulses detected across the frequency range of approximately 100\,MHz \citep{2021ApJ...911L...3P} to 8\,GHz \citep{2018ApJ...863....2G}, remain one of the rarest astronomical events observed at cosmological distances. So far, 843 FRBs have been reported, out of which 57 exhibit repeating behavior, and 115 have been precisely localized to their respective host galaxies\footnote{as of May 2025}. The maximum redshift observed for these host galaxies extends up to $z\approx1.35$ \citep{2025NatAs.tmp..131C}. The characteristically high flux densities and millisecond-scale durations of FRBs make them exceptional probes of host environments, which are predominantly extragalactic. A singular exception is FRB\,20200428A, which has been unambiguously associated with the Galactic magnetar SGR\,1935+2154 \citep{2020PASP..132c4202B,2020Natur.587...59B,2020Natur.587...54C}, providing direct observational support for the magnetar progenitor hypothesis. However, the absence of multi-wavelength counterparts, gravitational waves, or neutrino emissions coincident with extragalactic FRBs leaves their dominant physical mechanisms and progenitor populations unresolved \citep{2023MNRAS.520.3742K}, although theoretical models predominantly favor magnetars and compact object mergers.

A ubiquitous feature of all FRBs is the presence of a dispersion sweep in the frequency-time domain, corresponding to a propagation-induced time delay through ionized plasma along the line of sight. The extent of this dispersion is quantitatively expressed by the dispersion measure (DM), which represents the integrated free electron column number density encountered by the signal. For the majority of FRBs, the observed DM substantially exceeds the predicted Galactic contribution along the line of sight, indicating significant contributions from the intergalactic medium (IGM) and host galaxy. Consequently, FRBs have emerged as novel tools for probing IGM and by extension, constraining cosmological parameters. In particular, recent studies have utilized localized FRBs with measured redshifts and DMs to estimate the Hubble constant ($H_0$) under the standard $\Lambda$ cold dark matter ($\Lambda$CDM) framework \citep{2022MNRAS.511..662H,2022MNRAS.516.4862J,2022MNRAS.515L...1W,2025JCAP...01..018F,2025PDU....4801926K}.

A persistent challenge in modern cosmology is the Hubble tension: an approximately 5$\sigma$ discrepancy between late-Universe measurements of $H_0$ (primarily derived from Type Ia supernovae (SNe\,Ia) calibrated with Cepheid variables) and early-Universe inferences from cosmic microwave background (CMB) data assuming $\Lambda$CDM cosmology. Various strategies have been proposed to resolve this tension, including the adoption of alternative cosmological models and the utilization of novel astrophysical probes, among which FRBs have recently emerged as promising candidates (see the recent comprehensive review by \citealt{2025PDU....4901965D}). However, the limited number of well-localized FRBs and substantial uncertainties in modeling host galaxy contributions and the ionized baryon fraction in the IGM along different sight-lines currently prevent robust constraints on $H_0$ from FRB observations alone.

In this article, we employ a dual contemporary analytical approach, integrating machine learning (ML) algorithms with Bayesian inference, to analyze a dataset of well-localized FRBs from multiple telescopes. Our objective is to rigorously examine the internal consistency of the $\Lambda$CDM model at late cosmic epochs, where its validity is often presumed. We particularly investigate whether $H_0$ remains invariant across redshifts as predicted by $\Lambda$CDM, or if any detected deviations indicate a potential failure of this fundamental assumption. Our results reveal a statistically significant redshift dependence of $H_0$ when inferred from FRB data within the $\Lambda$CDM context while employing independent priors on baryon density parameters $\Omega_\mathrm{b}$ or $\Omega_\mathrm{b}h^2$, which, however, can be removed either by combining the priors or by considering a more generalized cosmological framework featuring a time-varying dark energy equation of state model. These findings suggest that the standard $\Lambda$CDM model may be insufficient to fully capture the complexity of current observational data and imply that the Hubble tension may reflect an intrinsic limitation of the $\Lambda$CDM model or inadequate knowledge of priors of other cosmological parameters, rather than solely arising from observational systematic uncertainties.

This article is structured as follows. In Section~\ref{Sec2}, we review the fundamental concepts of DM for FRBs and reconstruct the DM--redshift relation using a ML algorithm. In Section~\ref{Sec3}, we examine the invariance of $H_0$ within the $\Lambda$CDM framework by employing the reconstructed DM--redshift relation, followed by an independent analysis using Bayesian methods. We further validate these findings in an extended cosmological framework. In Section~\ref{Sec4}, we discuss our results, demonstrating that certain extended gravity models exhibit greater consistency than the $\Lambda$CDM model. Finally, we present our concluding remarks in Section~\ref{Sec5}.

\section{Decomposing dispersion measure of FRBs}\label{Sec2}

The observed DM of a FRB is attributed to contributions from four distinct astrophysical regions along the line of sight: the MW interstellar medium ($\mathrm{DM}_\mathrm{MW}$), Galactic halo ($\mathrm{DM}_\mathrm{Halo}$), IGM ($\mathrm{DM}_\mathrm{IGM}$), and host galaxy ($\mathrm{DM}_\mathrm{Host}$). For a source located at redshift $z_\mathrm{s}$, the cumulative observed DM can be expressed as
\begin{align}\label{Eq: DM}
    \mathrm{DM} = \mathrm{DM}_\mathrm{MW} + \mathrm{DM}_\mathrm{Halo} + \mathrm{DM}_\mathrm{IGM}(z_\mathrm{s}) + \frac{\mathrm{DM}_\mathrm{Host}}{1+z_\mathrm{s}}.
\end{align}
Among these four components, the Galactic contribution $\mathrm{DM}_\mathrm{MW}$ is the most robustly constrained owing to well-characterized models of the Galactic free electron density \citep{2002astro.ph..7156C,2017ApJ...835...29Y}. Accordingly, we define the residual DM as $\mathrm{DM}' = \mathrm{DM} - \mathrm{DM}_\mathrm{MW}$, which we compute for each of the localized FRBs. These values are plotted as a function of $z_\mathrm{s}$ in Fig.~\ref{Fig: DM_z_ML}, with the complete dataset provided in Appendix~\ref{Appendix1}. The other two factors $\mathrm{DM}_\mathrm{IGM}$ and $\mathrm{DM}_\mathrm{Host}$ are largely unknown, especially $\mathrm{DM}_\mathrm{Host}$ as the host galaxy morphology and the location of the FRB within it vary widely. Thus, $\mathrm{DM}_\mathrm{Host}$ can differ significantly from event to event, introducing substantial scatter into $\mathrm{DM}'$. As a result, we deal with the average values in these quantities subsequently.

\begin{figure}[htpb]
    \centering
    \includegraphics[scale=0.53]{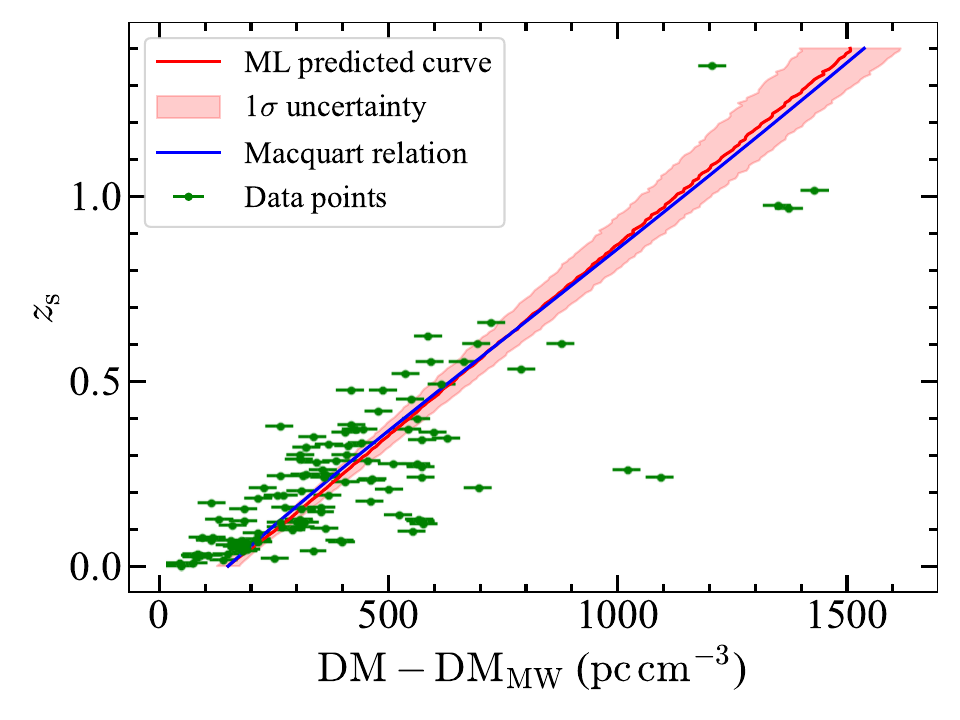}
    \caption{Observed and predicted $\mathrm{DM}'-z_\mathrm{s}$ relation. The green points represent measured redshifts $z_\mathrm{s}$ values for the localized FRBs plotted against the corresponding $\mathrm{DM}' = \mathrm{DM} - \mathrm{DM}_\mathrm{MW}$ including observational uncertainties. The red curve shows the ML prediction, with the shaded region denoting the associated 1$\sigma$ uncertainty. For comparison, we plot the Macquart relation combined with mean host and halo contribution in blue line with $H_0 = 73 \rm\,km\,s^{-1}\,Mpc^{-1}$ and $f_\mathrm{IGM}=0.85$ under $\Lambda$CDM cosmology.}
    \label{Fig: DM_z_ML}
\end{figure}

For the estimation of $\mathrm{DM}_\mathrm{MW}$, we adopt the NE2001 electron density model \citep{2002astro.ph..7156C}, which provides direction-dependent values. The associated uncertainties in $\mathrm{DM}'$ are computed using the error propagation relation $\sigma^2 = \sigma_{\mathrm{obs}}^2 + \sigma_{\mathrm{MW}}^2$, where $\sigma_\mathrm{obs}$ represents the measurement uncertainty in the observed DM and $\sigma_\mathrm{MW}$ accounts for the modeling uncertainty in $\mathrm{DM_\mathrm{MW}}$, which we adopt $\sigma_\mathrm{MW}=30\rm\,pc\,cm^{-3}$ following \cite{2005AJ....129.1993M}. Note that, in general, $\sigma_\mathrm{obs}\ll \sigma_\mathrm{MW}$ due to high-precision determination of $\mathrm{DM}$ from the signal. Thus, the error bars in $\mathrm{DM}'$ is dominated by the contribution from $\sigma_\mathrm{MW}$. To obtain $\mathrm{DM}_\mathrm{IGM}$, host and halo contributions need to be subtracted from $\mathrm{DM}'$, which will result in enhancing the error bars. Owing to the high-precision localization of the sources, the uncertainties in $z_\mathrm{s}$ are negligible in the present analysis.


\subsection{Reconstructing dispersion measure with machine learning algorithm}

We begin our analysis by applying a ML framework utilizing an artificial neural network (ANN) to model the empirical $\mathrm{DM}'-z_\mathrm{s}$ relation for localized FRBs. Our ANN architecture comprises of an input layer, two hidden layers (each containing 100 neurons), and an output layer. Nonlinearities in the model are introduced using the rectified linear unit (ReLU) activation function, and optimization is performed using the Adam algorithm within the \texttt{PyTorch} library. The resulting ANN-predicted $\mathrm{DM}'-z_\mathrm{s}$ trend is shown as the red solid line in Fig.~\ref{Fig: DM_z_ML}, with the shaded region indicating the 1$\sigma$ uncertainty bounds, which are estimated using a Bayesian uncertainty estimation integrated into the ML pipeline. Full details of the network architecture, hyperparameters, and training methodology are provided in Appendix~\ref{Appendix2}. The intercept of the curve at $\mathrm{DM}'\approx150 \rm\,pc\,cm^{-3}$ at $z_\mathrm{s}=0$ corresponds to the combined contribution from the Galactic halo and host galaxy, $\mathrm{DM}_\mathrm{Halo}+\mathrm{DM}_\mathrm{Host}$. While the individual contributions of these components are not precisely known for each FRB, previous studies have constrained the halo contribution to lie within the range $\mathrm{DM}_\mathrm{Halo}\approx 50-80\rm\,pc\,cm^{-3}$, independent of the contribution from the Galactic interstellar medium \citep{2019MNRAS.485..648P}. Adopting a mean halo contribution $\langle\mathrm{DM}_\mathrm{Halo}\rangle = 65\rm\,pc\,cm^{-3}$, we infer an average host galaxy contribution of $\langle\mathrm{DM}_\mathrm{Host}\rangle \approx 85\rm\,pc\,cm^{-3}$. Subtracting these values from the ML-predicted $\mathrm{DM}'$, we isolate the average IGM contribution $\langle\mathrm{DM}_\mathrm{IGM}\rangle$ as a function of redshift.

\section{Testing the Hubble Constant with FRB data}\label{Sec3}

Theoretically, $\langle\mathrm{DM}_\mathrm{IGM}\rangle$ can be computed by integrating average electron number density along the line of sight $\langle n_\mathrm{e}\rangle$ with luminosity distance $l$, i.e., $\langle\mathrm{DM}_\mathrm{IGM}\rangle = \int \langle n_\mathrm{e}\rangle (1+z)^{-1}\dd{l}$~\citep{2014ApJ...783L..35D}. Denoting $c$ as the speed of light, $G$ the gravitational constant, $m_\mathrm{p}$ the proton mass, $f_\mathrm{IGM}$ the baryon fraction in the IGM, and $\chi(z)$ the ionization fraction accounting for hydrogen and helium, $\langle\mathrm{DM}_\mathrm{IGM}\rangle$ can be explicitly written as
\citep{2020Natur.581..391M}
\begin{align}\label{Eq: Macquart}
    \langle \mathrm{DM}_\mathrm{IGM}(z_\mathrm{s})\rangle = \frac{3c \Omega_\mathrm{b} H_0^2}{8\pi G m_\mathrm{p}} \int_{0}^{z_\mathrm{s}} \frac{f_\mathrm{IGM}(z)\chi(z)(1 + z)}{H(z)} \dd{z},
\end{align}
where $H(z)$ is the Hubble parameter. Under the standard $\Lambda$CDM framework, neglecting radiation and curvature components, it is given by $H(z) = H_0E(z) = H_0\sqrt{\Omega_\mathrm{m} \left(1+z\right)^3 + \Omega_\Lambda}$ with $\Omega_\mathrm{m}$ and $\Omega_\Lambda$ respectively representing the present-day matter and dark energy density parameters with $\Omega_\mathrm{m} + \Omega_\Lambda = 1$. At low redshifts, as the Macquart relation is nearly linear, recovering a similar relation in our ML-predicted curve is expected. Moreover, from Fig.~\ref{Fig: DM_z_ML}, the network’s inherent tendency to capture the mean relation rather than the full variance is evident. As the intrinsic scatter in $\mathrm{DM}$ components cannot be removed, we focus on a global trend rather than fitting individual fluctuations. It is also scientifically appealing when we need to deal with $\langle \mathrm{DM}_\mathrm{IGM}\rangle$ from Eq.~\eqref{Eq: Macquart}.

\subsection{Machine learning based inference}

\begin{figure}[htpb]
    \centering
    \includegraphics[scale=0.53]{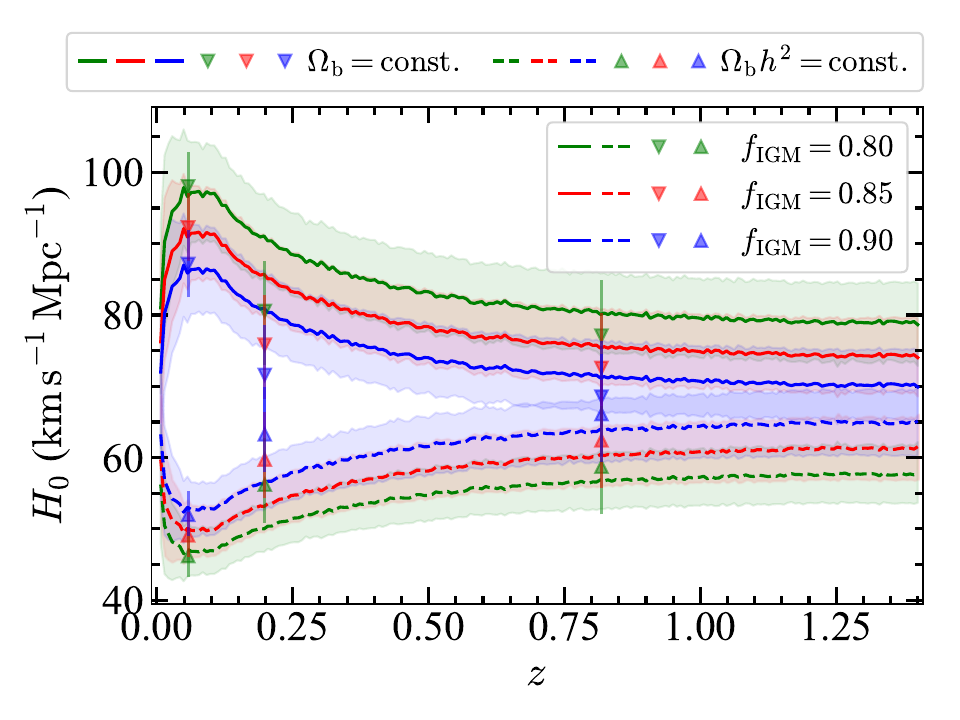}
    \caption{Variation of $H_0$ with respect to redshift for different $f_\mathrm{IGM}$. Solid and dashed lines denote the results obtained using the trained ML model, while scatter points represent the corresponding values derived from the Bayesian analysis.}
    \label{Fig: Hubble}
\end{figure}

For the baseline cosmological parameters, we adopt values from the Dark Energy Survey (DES) Data Release 1, $\Omega_\mathrm{b} = 0.0487^{+0.0005}_{-0.0004}$ and $\Omega_\mathrm{m} = 0.306^{+0.004}_{-0.005}$ \citep{2022PhRvD.105b3520A} with a constant $f_\mathrm{IGM}$ value of $f_\mathrm{IGM} = 0.85\pm0.05$ following \cite{2022ApJ...931...88C}, which is justified given the lack of significant redshift evolution up to $z\sim1$. Using these inputs and the ML-inferred $\langle\mathrm{DM}_\mathrm{IGM}\rangle$, we invert Eq.~\eqref{Eq: Macquart} to estimate the redshift evolution of $H_0$. Our analysis reveals a statistically significant decrease in the inferred $H_0$ with increasing $z_\mathrm{s}$ as illustrated by solid lines in Fig.~\ref{Fig: Hubble} for different $f_\mathrm{IGM}$. Notably, the rate of decrease of $H_0$ appears to diminish at higher redshifts.

Conversely, adopting cosmological parameters from DESI+CMB Data Release 1 ($\Omega_\mathrm{b} h^2 = 0.02218 \pm 0.00055$ and $\Omega_\mathrm{m} = 0.3069 \pm 0.0050$, where $h \equiv H_0 / 100\rm\, \mathrm{km\,s^{-1}\,Mpc^{-1}}$) following \cite{2025JCAP...02..021A} results in a statistically significant increase in $H_0$ with $z_\mathrm{s}$, depicted as dashed lines in the same figure. The opposing trends with the previous case can be interpreted as follows. From Eq.~\eqref{Eq: Macquart}, the independent quantity is $\Omega_\mathrm{b}H_0$, and FRB data indicate that this product decreases with increasing $z$. Consequently, for a fixed $\Omega_\mathrm{b}$, $H_0 = (\Omega_\mathrm{b}H_0)/\Omega_\mathrm{b}$ and it must decrease with $z$. On the other hand, imposing a fixed $\Omega_\mathrm{b} h^2$ means $H_0 = 10^4\Omega_\mathrm{b} h^2/(\Omega_\mathrm{b}H_0)$ and thus $H_0$ increases with $z$. Hence, both scenarios complement each other.

The decreasing trend of $\Omega_\mathrm{b}H_0$ conflict with the $\Lambda$CDM requirement and as a result, both scenarios independently conflict with the $\Lambda$CDM requirement of an invariant $H_0$, suggesting a fundamental inconsistency between the observed $\mathrm{DM}'-z_\mathrm{s}$ relation and standard cosmological framework. This redshift-dependent tension implies either unaccounted systematics in FRB-derived DMs or a breakdown of $\Lambda$CDM assumptions, even at late cosmic times. Hence, the observed Hubble tension may be an artifact of $\Lambda$CDM model, rather than indicating a genuine physical phenomenon. Furthermore, it is important to mention that using the Macquart relation of Eq.~\eqref{Eq: Macquart}, FRB data alone can constrain only one cosmological parameter, or at best, some specific combinations of these parameters. It is not possible to extract the individual values of all cosmological parameters solely from FRB observations. To break these degeneracies, FRB data must be combined with independent measurements from other cosmological probes, as shown using a large number of mock FRBs by \cite{2018ApJ...856...65W}.


\subsection{Bayesian inference}
In our second methodology, in place of ML algorithm, we implement a Bayesian hierarchical framework to independently test the redshift dependence of $H_0$. To observe any changes in $H_0$ with $z$, the dataset is divided into discrete redshift bins. Ideally, a finer binning scheme would be preferred to improve resolution; however, due to the limited number of localized FRBs, we restrict our analysis to three redshift bins $[0-0.115]$, $[0.115-0.282]$, and $[0.282-1.4]$, each respectively containing 36, 40, and 38 samples, with $H_0$ treated as distinct parameters $H_{01}$, $H_{02}$, and $H_{03}$ within each bin. To accommodate potential redshift-dependent $H_0$, we now rewrite the Macquart relation under $\Lambda$CDM model as follows
\begin{align}\label{Eq: Macquart2}
    \langle \mathrm{DM}_\mathrm{IGM}(z_\mathrm{s})\rangle &= \frac{3c \Omega_\mathrm{b} H_{01}^2}{8\pi G m_\mathrm{p}} \int_{0}^{z_1} \frac{f_\mathrm{IGM}(z)\chi(z)(1 + z)}{H_{01}\sqrt{\Omega_\mathrm{m} \left(1+z\right)^3 + \Omega_\Lambda}} \dd{z} \nonumber\\
    &+ \frac{3c \Omega_\mathrm{b} H_{02}^2}{8\pi G m_\mathrm{p}} \int_{z_1}^{z_2} \frac{f_\mathrm{IGM}(z)\chi(z)(1 + z)}{H_{02}\sqrt{\Omega_\mathrm{m} \left(1+z\right)^3 + \Omega_\Lambda}} \dd{z} \nonumber\\
    &+ \frac{3c \Omega_\mathrm{b} H_{03}^2}{8\pi G m_\mathrm{p}} \int_{z_2}^{z_\mathrm{s}} \frac{f_\mathrm{IGM}(z)\chi(z)(1 + z)}{H_{03}\sqrt{\Omega_\mathrm{m} \left(1+z\right)^3 + \Omega_\Lambda}} \dd{z},
\end{align}
where $z_1$ and $z_2$ are the upper redshift boundaries of the first and second bins, respectively. For a source with $z_\mathrm{s}<z_1$, only the first integral contributes; while for $z_1<z_\mathrm{s}<z_2$, the first two terms are relevant; and for $z_\mathrm{s}>z_2$, all three integrals contribute to $\langle \mathrm{DM}_\mathrm{IGM}\rangle$. We now use the most general likelihood function for evaluating $H_{01}$, $H_{02}$, and $H_{03}$ as follows
\begin{align}\label{Eq: Likelihood3}
    \mathcal{L} = \prod_{i=1}^{N_\mathrm{FRB}} P_i\left(\mathrm{DM}'_{i} \mid z_{\mathrm{s},i}\right),
\end{align}
where the probability for each individual FRB is given by \citep{2025PDU....4801926K}
\begin{align}
     P_i\left(\mathrm{DM}'_{i} \mid z_{\mathrm{s},i}\right) &= \int_0^{\mathrm{DM}'_{i}} \int_0^{\mathrm{DM}'_{i} - \mathrm{DM}_\mathrm{Halo}}  P_\mathrm{Halo}\left(\mathrm{DM}_{\mathrm{Halo}}\right) \nonumber\\ &\times P_\mathrm{IGM}\left(\mathrm{DM}'_{i} - \frac{\mathrm{DM}_\mathrm{Host}}{1+z_{\mathrm{s},i}} - \mathrm{DM}_\mathrm{Halo} \right) \nonumber\\ &\times P_\mathrm{Host}\left(\frac{\mathrm{DM}_\mathrm{Host}}{1+z_{\mathrm{s},i}}\right) \dd{\mathrm{DM}_\mathrm{Halo}} \dd{\mathrm{DM}_\mathrm{Host}}.
\end{align}
Here, $P_\mathrm{Halo}$, $P_\mathrm{IGM}$, and $P_\mathrm{Host}$ represent the probability density functions for the halo, IGM, and host galaxy DM contributions, respectively. The functional forms and parameterizations of these distributions, as well as the details of the Bayesian inference implementation, are provided in Appendix~\ref{Appendix3}.

The inferred $H_{01}$, $H_{02}$, and $H_{03}$ along with their associated uncertainties, assuming flat priors, are presented in Fig.~\ref{Fig: Hubble} for different $f_\mathrm{IGM}$ values. Consistent with the results obtained from the ML analysis, the Bayesian approach also reveals a statistically significant evolution of $H_0$ with redshift. The opposing trend of this evolution again depends on whether $\Omega_\mathrm{b}h^2$ or $\Omega_\mathrm{b}$ is held fixed. Taken together, the results from both ML and Bayesian frameworks provide compelling evidence that the assumption of a constant $H_0$ in $\Lambda$CDM model does not hold across all redshifts. These findings verify the aforementioned outcomes based on ML algorithm and point toward a potential breakdown of the $\Lambda$CDM framework, even during the late-time epoch.

 From Fig.~\ref{Fig: Hubble}, we notice that FRBs at higher $z_\mathrm{s}$ tend to favor lower values of $\Omega_\mathrm{b}H_0$ compared to those at lower $z_\mathrm{s}$. Consequently, imposing independent priors from either DES or DESI+CMB measurements introduces opposite trends in the inferred evolution of $H_0$ with redshift. It was previously shown that the results inferred from these two measurements are in tension~\citep{2025MNRAS.542L..24O}
and to account for this discrepancy, we explore a joint prior that combines both sets of constraints.

\begin{figure}[htpb]
    \centering
    \includegraphics[scale=0.74]{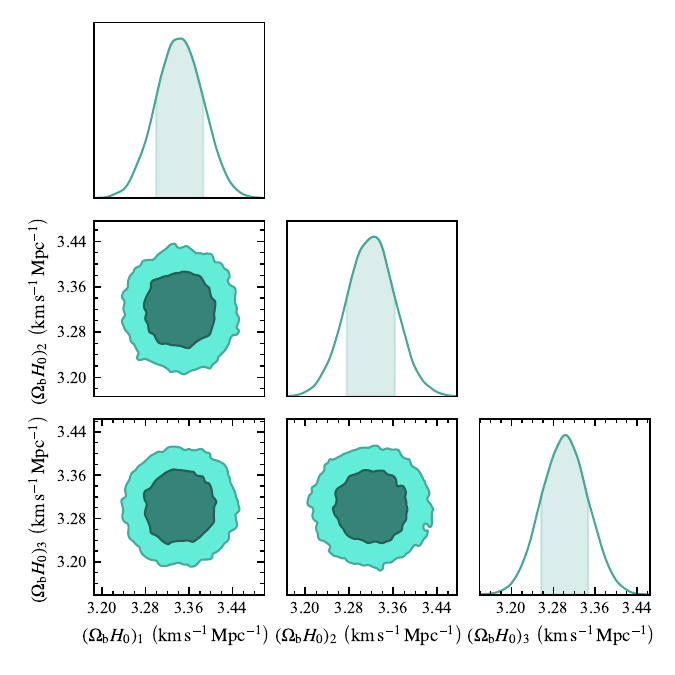}
    \caption{Posterior distribution of $\Omega_\mathrm{b}H_0$ across three redshift bins, derived using the joint prior that combines constraints from DES and DESI+CMB.}
    \label{Fig: Combine_prior}
\end{figure}

We assume normal prior distributions for $\Omega_\mathrm{b}$ with mean $\mu_{\Omega_\mathrm{b}}$ and variance $\sigma_{\Omega_\mathrm{b}}^2$, and for $\Omega_\mathrm{b}h^2$ with mean $\mu_{\Omega_\mathrm{b}h^2}$ and variance $\sigma_{\Omega_\mathrm{b}h^2}^2$. The relation between these quantities and the parameter of interest is given by $\Omega_\mathrm{b}H_0 = 100 \sqrt{\Omega_\mathrm{b}(\Omega_\mathrm{b}h^2)}$. As the uncertainties in both priors are relatively small, the resulting combined prior on $\Omega_\mathrm{b}H_0$ can be well approximated by a normal distribution $\mathcal{N}(\mu,\sigma^2)$ with mean $\mu = 100 \sqrt{\mu_{\Omega_\mathrm{b}}\mu_{\Omega_\mathrm{b}h^2}}$ and variance $\sigma^2 = [(\sigma_{\Omega_\mathrm{b}}^2/\Omega_\mathrm{b}^2) + (\sigma_{\Omega_\mathrm{b}h^2}^2/(\Omega_\mathrm{b}h^2)^2)]/4$. The corresponding posterior probability can be expressed as
\begin{align}
    \mathcal{L}(\Omega_\mathrm{b}H_0\mid\text{data}) \propto \mathcal{L}(\text{data}\mid\Omega_\mathrm{b}H_0) \mathcal{P}(\mathcal{N}(\mu,\sigma^2)),
\end{align}
where $\mathcal{L}(\text{data}\mid\Omega_\mathrm{b}H_0)$ is equivalent to Eq.~\eqref{Eq: Likelihood3}. Using this new combined prior, we compute $\Omega_\mathrm{b}H_0$ for the aforementioned three bins. For $f_\mathrm{IGM}=0.85$, the resulting value of $\Omega_\mathrm{b}H_0$ in the units of $\rm km\,s^{-1}\,Mpc^{-1}$ under 1$\sigma$ uncertainty in each bin respectively turns out to be $3.343\pm0.043$, $3.320\pm0.044$, and $3.302\pm0.045$, as shown in Fig.~\ref{Fig: Combine_prior}. Although a mild decreasing trend with redshift is visible, it is not statistically significant. The corresponding $H_0$ in the units of $\rm km\,s^{-1}\,Mpc^{-1}$ is $68.649\pm1.130$, $68.173\pm1.127$, and $67.803\pm1.124$ substituting $\Omega_\mathrm{b}$ from DES; whereas $66.348\pm1.853$, $66.807\pm1.869$, $67.171\pm1.881$ substituting $\Omega_\mathrm{b}h^2$ from DESI+CMB. Hence, the joint prior yields an effectively redshift-invariant estimate of $H_0$, mitigating the alteration trends that arise when DES and DESI+CMB priors are applied separately.


\subsection{Alternate cosmology model}
\begin{figure}[htpb]
    \centering
    \subfigure[Using ML-predicted $\mathrm{DM}'-z_\mathrm{s}$ curve.]{\includegraphics[scale=0.5]{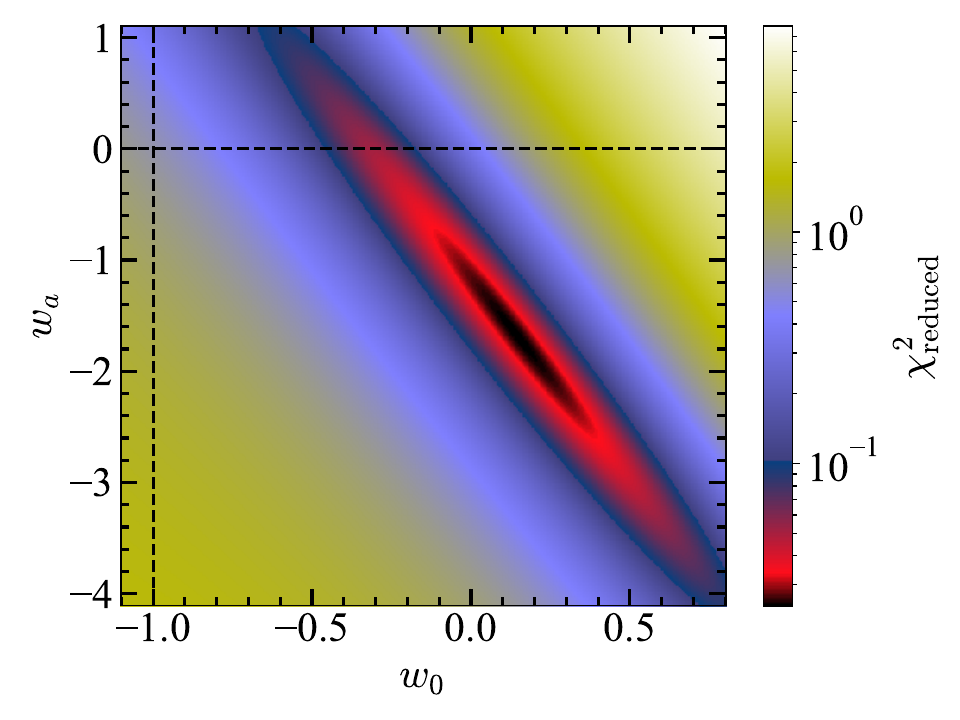}}    
    \subfigure[Using Bayesian analysis on binned data.]{\includegraphics[scale=0.5]{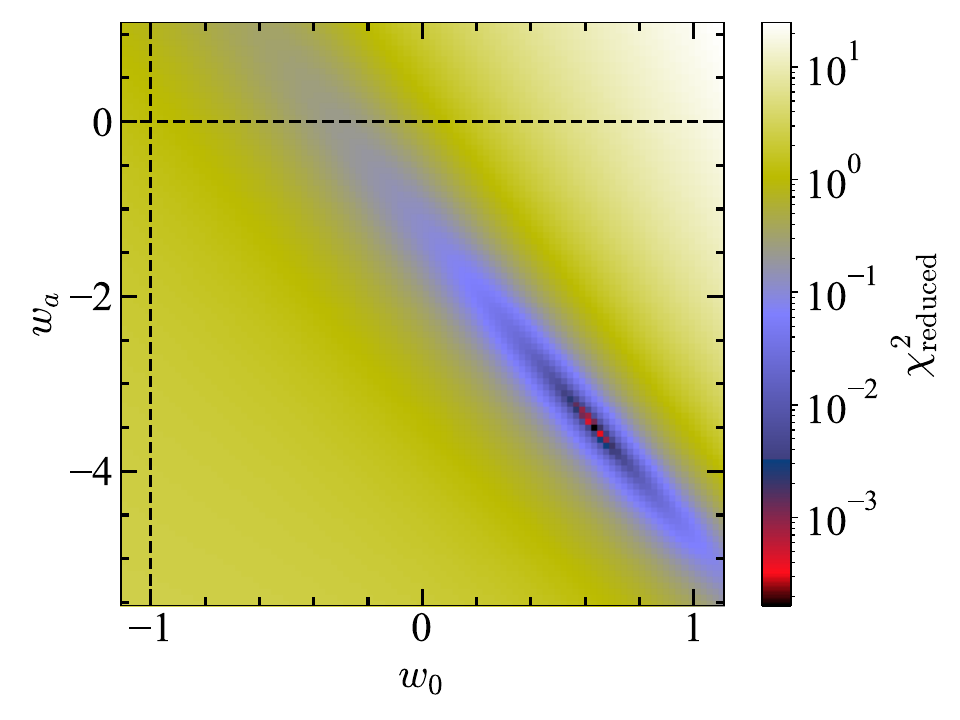}}    
    \caption{Reduced $\chi^2$ value for different combinations of ($w_0,w_a$). This contour plot is independent of value of $f_\mathrm{IGM}$ and of the choice to fix $\Omega_\mathrm{b}$ or $\Omega_\mathrm{b}h^2$. Different combinations of ($w_0,w_a$) yield $\chi^2_\mathrm{reduced}\ll1$ supporting the statistical consistency of a redshift-independent $H_0$ within these models.}
    \label{Fig: Diff_dynamical}
\end{figure}

It remains uncertain which of the two aforementioned observational priors provides a more accurate description of the true cosmological parameters. Therefore, we investigate whether each of these priors can independently recover a redshift-invariant $\Omega_\mathrm{b}H_0$ (thereby redshift-invariant $H_0$) when tested under an alternative cosmological framework. In particular, we extend our analysis to a widely studied alternative framework involving a dynamical dark energy equation of state. Specifically, we consider the Chevallier–Polarski–Linder (CPL) parameterization \citep{2001IJMPD..10..213C,2003PhRvL..90i1301L}, in which the dark energy equation of state evolves with redshift as $w(z) = w_0 + w_a z/(1+z)$ with $w_0$ and $w_a$ being dimensionless model parameters. This leads to a modification of the Hubble parameter as $H(z) = H_0\sqrt{\Omega_\mathrm{m} \left(1+z\right)^3 + \Omega_\Lambda f(z)}$ with the dark energy evolution factor is given by $f(z) = (1+z)^{3\left(1+w_0+w_a\right)} \exp{-3 w_a z/(1+z)}$ \citep{2005PhRvD..72j3503J}. In the CPL parametrization, the dynamical term primarily affects the dark energy evolution at intermediate and high redshifts, while near the present epoch it effectively reduces to a $\Lambda$CDM model. Using this generalized form of $H(z)$, we compute the redshift-dependent values of $H_0$ for various combinations of ($w_0,w_a$) using the aforementioned ML algorithm.
To quantitatively evaluate whether a given model supports a constant $H_0$, we introduce the reduced $\chi^2$ function as
\begin{align}
    \chi^2_\mathrm{reduced} = \frac{1}{N-1} \sum_{i=1}^{N} \left(\frac{H_{0,i}-H_{0,\mathrm{best}}}{\sigma_{H_{0,i}}}\right)^2,
\end{align}
where $H_{0,\mathrm{best}}$ denotes the best-fit constant value across $N$ points chosen from the curve and $H_{0,i}$ is the inferred $H_0$ value for the $i^\mathrm{th}$ data point with $\sigma_{H_{0,i}}$ being its 1$\sigma$ uncertainty. The resulting values of $\chi^2_\mathrm{reduced}$ for different combinations of $w_0$ and $w_a$ are presented in Fig.~\ref{Fig: Diff_dynamical}. Note that this plot is unaffected by the choice of $f_\mathrm{IGM}$ and is valid for both cases of fixed $\Omega_\mathrm{b}$ or $\Omega_\mathrm{b}h^2$.

The differences observed in both panels of Fig.~\ref{Fig: Diff_dynamical} arise predominantly from methodological disparities. Specifically, the binning approach employs only three bins, each containing approximately 38 FRBs. With more FRBs localized in the future, we may achieve better precision on parameters with finer binning. Nevertheless, it is evident from the panels of Fig.~\ref{Fig: Diff_dynamical} that the standard $\Lambda$CDM model ($w_0=-1$ and $w_a=0$) yields $\chi^2_\mathrm{reduced} > 1$, indicating a poor fit to the hypothesis of constant $H_0$ as previously discussed. In contrast, several alternative parameter combinations yield $\chi^2_\mathrm{reduced}\ll1$, thereby providing statistically robust support for redshift-independent $\Omega_\mathrm{b}H_0$, thereby $H_0$, which are characterized by $w_0>-1$ and $w_a<0$, aligning with the majority of results reported in prior literature \citep{2022PhRvD.105b3520A,2025JCAP...02..021A}.  The exact value of $\Omega_\mathrm{b}H_0$ depends on the specific values of $f_\mathrm{IGM}$ as $\Omega_\mathrm{b}H_0\propto f_\mathrm{IGM}^{-1}$ from the Macquart relation of Eq.~\eqref{Eq: Macquart}. For $f_\mathrm{IGM} = 0.9$, we obtain $\Omega_\mathrm{b}H_0\ = 4.282\pm0.294\rm\,km\,s^{-1}\,Mpc^{-1}$. The corresponding $H_0$ is $87.927\pm6.080 \rm\, km\,s^{-1}\,Mpc^{-1}$ substituting $\Omega_\mathrm{b}$ from DES and $51.798\pm 3.781\rm\, km\,s^{-1}\,Mpc^{-1}$ substituting $\Omega_\mathrm{b}h^2$ from DESI+CMB. This behavior is consistent with the results shown in Fig.~7 of \cite{2025PDU....4801926K}, where using Bayesian analysis it was shown that $H_0$ tends to increase in the $w_0w_a$CDM model with $w_0>-1$ and $w_a<0$ for fixed $\Omega_\mathrm{b}$.


\section{Discussion}\label{Sec4}

In the present study, we have analyzed a sample of well-localized FRBs using a dual methodology that combines ML and Bayesian hierarchical modeling to test the redshift invariance of $H_0$ under the $\Lambda$CDM model. Both methodologies independently reveal a statistically significant redshift-dependent $H_0$, depending on the parameter priors, directly contradicting $\Lambda$CDM’s foundational assumption of a constant $H_0$. Similar redshift-dependent trends in $H_0$ under $\Lambda$CDM model have been reported in some previous studies. The H0LiCOW collaboration, employing six gravitationally lensed quasars with accurately measured time delays, identified a decreasing trend in $H_0$ with redshift \citep{2020MNRAS.498.1420W}. Subsequent analyses using SN\,Ia datasets also revealed analogous behavior, either manifesting as a decline in $H_0$ or an increasing trend in $\Omega_\mathrm{m}$ with redshift \citep{2020PhRvD.102j3525K,2020PhRvD.102b3520K,2021PhRvD.103l1302C,2021ApJ...912..150D,2023A&A...674A..45J,2025JHEAp..4800405D,2025JHEAp..45..290D,2025JCAP...03..026L,2025MNRAS.536.3232M}. This inverse trend of $\Omega_\mathrm{m}$ and $H_0$ may be attributable to the increasing dominance of dark energy at late cosmic times, which could suppress the observable correlations with $\Omega_\mathrm{m}$. Consequently, it was proposed that in the absence of significant statistical anomalies, local matter underdensities or modifications to general relativity may be critical factors underlying these discrepancies \citep{2020PhRvD.102j3525K}. Regardless of interpretation, the presence of these trends in diverse datasets needs attention, emphasizing their relevance in the context of the Hubble tension.

$H_0$ naturally arises as a proportionality constant when solving the Friedmann equation. More explicitly, it can be written as
    \begin{align}
        H(z) = H_0 \exp\left(\frac{3}{2} \int_0^z \frac{1+w(z')}{1+z'}\,dz'\right).
    \end{align}
Once a cosmological model is specified, the exponential term on the right-hand side is just a function of $z$, implying $H_0$ should not depend on $z$. If $H_0$ is truly a physical constant within a given cosmological model, the rate of change of the left-hand side and the exponential term on the right-hand side should exactly be the same. However, our result shows that within $\Lambda$CDM, $H_0$ tends to vary, suggesting a mismatch between these rates of change. To test whether this variation is intrinsic or model-dependent, we extended our analysis  by combining the priors on $\Omega_\mathrm{b}$ and $\Omega_\mathrm{b}h^2$ and found a redshift-independent $\Omega_\mathrm{b}H_0$. Moreover, we have examined our results for the $w_0w_a$CDM framework and found that for certain combinations of these parameters ($w_0>-1$ and $w_a<0$), the variation of $H_0$ can be removed  even without combining the priors. Hence, these findings,  along with the outcomes based on previous studies using SNe\,Ia, highlight a potential fundamental discrepancy in the standard cosmological framework, and thus the Hubble tension may infer how sensitive cosmological inferences are to prior tensions and the limited constraining power of FRB data~\citep{2023Univ....9..393V}.

Unlike previous studies that relied on only discrete binning, our ML-based framework enables a continuous reconstruction of $H_0(z)$, capturing fine-scale variations across the full redshift range. Our results underscore that while $H_0$ varies with redshift, the rate of change flattens at higher redshifts, indicating that the Hubble tension may be an artifact of the $\Lambda$CDM model or inadequate prior information. However, due to limited FRB data beyond $z\sim1$, extrapolation beyond this limit is not reliable, as the ML algorithm would not be accurate in that range. In general, ML algorithm benefits from larger training samples, which, however, is not possible as we are limited by a relatively small data set. Importantly, the observed redshift-dependent tension persists irrespective of the dataset or technique employed, emphasizing its robustness. Another minor limitation of this analysis is its dependence on the mean distributions of IGM and host DM. This methodology may inadequately represent cases where the observed DM is anomalously high, despite the source being located at a relatively low redshift. Additionally, the contribution from the host galaxy halo may differ significantly from that of the Milky Way halo, and its contribution could change over time. Nevertheless, given the current limitations of observational facilities in precisely determining individual DM components for each FRB, the use of mean values remains the most reliable and practical approach for this analysis. Extending the analysis to the dynamical dark energy model of $w_0w_a$CDM, we identify specific parameterizations of $w_0$ and $w_a$ that restore the statistical consistency of a redshift-independent $H_0$. These results suggest that the apparent variation in $H_0$ is likely a manifestation of the known, unresolved tensions among existing cosmological datasets, an effect that is further amplified by the relatively weak constraining power of the FRB data or may arise from unaccounted dark energy dynamics rather than observational biases.

\section{Conclusion}\label{Sec5}

These findings suggest that the Hubble tension may originate from the limitations inherent to the $\Lambda$CDM model, particularly its treatment of dark energy as a cosmological constant, rather than from systematic observational errors, advocating for the exploration of more general cosmological models even in the late Universe. Thus the expansion history of the Universe might not be fully captured by $\Lambda$CDM framework with a static cosmological constant, and hence, tackling problems like the Hubble tension  might require better understanding of dark energy dynamics and the fundamental physics governing cosmic acceleration.

\begin{acknowledgements}
The authors thank the anonymous reviewer for their insightful comments, which have helped to improve the quality of this manuscript, particularly for the suggestion to combine the priors from different cosmological datasets. S.K. gratefully acknowledges the support and hospitality from ICTP, Italy during a research visit through the Associates Programme. S.K. and T.B. gratefully acknowledge funding from the National Science Centre, Poland (grant no. 2023/49/B/ST9/02777). A.U. and Y.M. are supported by the National Key R\&D Program of China (grant no.\,2023YFE0101200), the National Natural Science Foundation of China (grant no.\,12273022), the Research Fund for Excellent International PhD Students (grant no.\,W2442004) and the Shanghai Municipality orientation program of Basic Research for International Scientists (grant no.\,22JC1410600). Simulations were performed on the TDLI-Astro cluster and Siyuan Mark-I at Shanghai Jiao Tong University.
\end{acknowledgements}


\bibliography{bibliography}{}
\bibliographystyle{aasjournal}



\appendix

\section{Localized FRB dataset}\label{Appendix1}

\begin{longtable}{|l|l|l|l|l|l|l|}
\caption{List of localized FRBs as of May 2025 in chronological order. $\mathrm{DM}_\mathrm{MW}$ is calculated based on NE2001 model.}
\label{Table: FRB}\\
\hline
Name & $\mathrm{DM}_\mathrm{obs}$ & $\mathrm{DM}_\mathrm{MW}$ & $z_\mathrm{s}$ & Repeater & Ref. \\
& $(\rm pc\,cm^{-3})$ & $(\rm pc\,cm^{-3})$  & & (Y/N) & \\
\hline
FRB\,20121102A & $557.0\pm2.0$ & 188.0 & 0.19273 & Y & \cite{2017ApJ...834L...7T}\\
FRB\,20171020A & $114.1\pm0.2$ & 38.0 & 0.0086 & N & \cite{2018ApJ...867L..10M} \\
FRB\,20180301A & $522.0\pm0.2$ & 152.0 & 0.3304 & Y & \cite{2019MNRAS.486.3636P}\\
FRB\,20180916B & $349.3\pm0.2$ & 200.0 & 0.0337 & Y & \cite{2020Natur.577..190M}\\
FRB\,20180924B & $361.42\pm0.06$ & 40.5 & 0.3214 & N & \cite{2019Sci...365..565B}\\
FRB\,20181112A & $589.27\pm0.03$ & 102.0 & 0.4755 & N & \cite{2019Sci...366..231P}\\
FRB\,20181220A & $209.4\pm0.1$ & 126.0 & 0.02746 & N & \cite{2024ApJ...971L..51B}\\
FRB\,20181223C & $112.5\pm0.1$ & 20.0 & 0.03024 & N & \cite{2024ApJ...971L..51B}\\
FRB\,20190102C & $363.6\pm0.3$ & 57.3 & 0.291 & N & \cite{2020Natur.581..391M}\\
FRB\,20190110C & $221.6\pm1.6$ & 37.1 & 0.12244 & Y & \cite{2024ApJ...961...99I}\\
FRB\,20190203A & $134.4\pm2.0$ & 20.3 & 0.17 & N & \cite{2025PASA...42...59T}\\
FRB\,20190208A & $580.01\pm0.26$ & 71.5 & 0.83 & Y & \cite{2024ApJ...977L...4H}\\
FRB\,20190303A & $222.4\pm0.7$ & 29.0 & 0.064 & Y & \cite{2023ApJ...950..134M}\\
FRB\,20190418A & $184.5\pm0.1$ & 71.0 & 0.07132 & N & \cite{2024ApJ...971L..51B}\\
FRB\,20190425A & $128.2\pm0.1$ & 49.0 & 0.03122 & N & \cite{2024ApJ...971L..51B}\\
FRB\,20190520B & $1204.7\pm10.0$ & 113 & 0.241 & Y & \cite{2022ApJ...931...87O}\\
FRB\,20190523A & $760.8\pm0.6$ & 37.0 & 0.66 & N & \cite{2019ApJ...886..135P}\\
FRB\,20190608B & $338.7\pm0.5$ & 37.2 & 0.1178 & N & \cite{2020ApJ...901..134S}\\
FRB\,20190611B & $321.4\pm0.2$ & 57.83 & 0.3778 & N & \cite{2020ApJ...903..152H}\\
FRB\,20190614D & $959.2\pm5.0$ & 83.5 & 0.6 & N & \cite{2020ApJ...899..161L}\\
FRB\,20190711A & $593.1\pm0.4$ & 56.4 & 0.522 & Y & \cite{2021MNRAS.500.2525K}\\
FRB\,20190714A & $504.0\pm2.0$ & 39.0 & 0.2365 & N & \cite{2020ApJ...903..152H}\\
FRB\,20191001A & $506.92\pm0.04$ & 44.7 & 0.234 & N & \cite{2022MNRAS.512L...1K}\\
FRB\,20191106C & $332.2\pm0.7$ & 25.0 & 0.10775 & Y & \cite{2024ApJ...961...99I}\\
FRB\,20191228A & $297.50\pm0.05$ & 33.0 & 0.2432 & N & \cite{2022AJ....163...69B}\\
FRB\,20200120E & 87.8 & 40.8 & 0.0008 & N & \cite{2025NatAs.tmp..131C}\\
FRB\,20200223B & $201.8\pm0.4$ & 45.6 & 0.0602 & Y & \cite{2024ApJ...961...99I}\\
FRB\,20200430A & $380.1\pm0.4$ & 27.0 & 0.16 & N & \cite{2020ApJ...903..152H}\\
FRB\,20200906A & $577.80\pm0.02$ & 36.0 & 0.3688 & N & \cite{2022AJ....163...69B}\\
FRB\,20201123A & $433.55\pm0.36$ & 251.93 & 0.05 & N & \cite{2022MNRAS.514.1961R}\\
FRB\,20201124A & $413.52\pm0.05$ & 123.2 & 0.098 & Y & \cite{2022MNRAS.513..982R}\\
FRB\,20210117A & $730.0\pm1.0$ & 34.4 & 0.2145 & N & \cite{2023ApJ...948...67B}\\
FRB\,20210320C & $384.8\pm0.3$ & 42.0 & 0.2797 & N & \cite{2022MNRAS.516.4862J}\\
FRB\,20210405I & $565.17\pm0.49$ & 516.1 & 0.066 & N & \cite{2024MNRAS.527.3659D}\\
FRB\,20210410D & $578.78\pm2.00$ & 56.2 & 0.1415 & N & \cite{2023MNRAS.524.2064C}\\
FRB\,20210603A & $500.147\pm0.004$ & 40.0 & 0.177 & N & \cite{2024NatAs...8.1429C}\\
FRB\,20210807D & $251.9\pm0.2$ & 121.2 & 0.12927 & N & \cite{2022MNRAS.516.4862J}\\
FRB\,20210924D & 737 & 45 & 0.6 & Y & \cite{2024MNRAS.534.3377C}\\
FRB\,20211127I & $234.83\pm0.08$ & 42.5 & 0.0469 & N & \cite{2022MNRAS.516.4862J}\\
FRB\,20211203C & $636.2\pm0.4$ & 63.0 & 0.3437 & N & \cite{2024arXiv240802083S}\\
FRB\,20211212A & $206.0\pm5.0$ & 27.1 & 0.0715 & N & \cite{2022MNRAS.516.4862J}\\
FRB\,20220105A & $583.0\pm2.0$ & 22.0 & 0.2785 & N & \cite{2024arXiv240802083S}\\
FRB\,20220204A & 612.2 & 50.7 & 0.4 & N & \cite{2024ApJ...968...50P}\\
FRB\,20220207C & $262.38\pm0.01$ & 79.3 & 0.04304 & N & \cite{2024ApJ...967...29L}\\
FRB\,20220208A & 437 & 101.6 & 0.351 & N & \cite{2024Natur.635...61S}\\
FRB\,20220307B & $499.27\pm0.06$ & 135.7 & 0.248123 & N & \cite{2024ApJ...967...29L}\\
FRB\,20220310F & $462.240\pm0.005$ & 45.4 & 0.477958 & N & \cite{2024ApJ...967...29L}\\
FRB\,20220319D & $110.95\pm0.02$ & 65.25 & 0.011 & N & \cite{2025AJ....169..330R}\\
FRB\,20220330D & 468.1 & 38.6 & 0.3714 & N & \cite{2024Natur.635...61S}\\
FRB\,20220418A & $623.25\pm0.01$ & 37.6 & 0.622 & N & \cite{2024ApJ...967...29L}\\
FRB\,20220501C & $449.5\pm0.2$ & 31.0 & 0.381 & N & \cite{2024arXiv240802083S}\\
FRB\,20220506D & $396.97\pm0.02$ & 89.1 & 0.30039 & N & \cite{2024ApJ...967...29L}\\
FRB\,20220509G & $269.53\pm0.02$ & 55.2 & 0.0894 & N & \cite{2024ApJ...967...29L}\\
FRB\,20220529A & $246.3\pm0.4$ & 30 & 0.1839 & Y & \cite{2025arXiv250304727L}\\
FRB\,20220610A & $1458.0\pm0.2$ & 31.0 & 1.017 & N & \cite{2024ApJ...963L..34G}\\
FRB\,20220725A & $290.4\pm0.3$ & 31.0 & 0.1926 & N & \cite{2024arXiv240802083S}\\
FRB\,20220726A & 686.55 & 89.5 & 0.361 & N & \cite{2025NatAs.tmp..131C}\\
FRB\,20220825A & $651.24\pm0.06$ & 79.7 & 0.241397 & N & \cite{2024ApJ...967...29L}\\
FRB\,20220831A & 1146.25 & 126.7 & 0.262 & N & \cite{2025NatAs.tmp..131C}\\
FRB\,20220912A & $219.46\pm0.04$ & 125.0 & 0.077 & Y & \cite{2023ApJ...949L...3R}\\
FRB\,20220914A & $631.28\pm0.04$ & 55.2 & 0.1139 & N & \cite{2024ApJ...967...29L}\\
FRB\,20220918A & $656.8\pm0.4$ & 41.0 & 0.491 & N & \cite{2024arXiv240802083S}\\
FRB\,20220920A & $314.99\pm0.01$ & 40.3 & 0.158239 & N & \cite{2024ApJ...967...29L}\\
FRB\,20221012A & $441.08\pm0.70$ & 54.4 & 0.284669 & N & \cite{2024ApJ...967...29L}\\
FRB\,20221027A & 452.5 & 47.2 & 0.229 & N & \cite{2024Natur.635...61S}\\
FRB\,20221029A & 1391.05 & 43.9 & 0.975 & N & \cite{2024Natur.635...61S}\\
FRB\,20221101B & 490.7 & 131.2 & 0.2395 & N & \cite{2024Natur.635...61S}\\
FRB\,20221106A & $343.8\pm0.8$ & 35.0 & 0.2044 & N & \cite{2024arXiv240802083S}\\
FRB\,20221113A & 411.4 & 91.7 & 0.2505 & N & \cite{2024Natur.635...61S}\\
FRB\,20221116A & 640.6 & 132.3 & 0.2764 & N & \cite{2024Natur.635...61S}\\
FRB\,20221219A & 706.7 & 44.4 & 0.554 & N & \cite{2024Natur.635...61S}\\
FRB\,20230124A & 590.6 & 38.5 & 0.094 & N & \cite{2024Natur.635...61S}\\
FRB\,20230203A & 420.1 & 67.3 & 0.1464 & N & \cite{2025ApJS..280....6C}\\
FRB\,20230216A & 828 & 38.5 & 0.531 & N & \cite{2024Natur.635...61S}\\
FRB\,20230222A & 706.1 & 134.2 & 0.1223 & N & \cite{2025ApJS..280....6C}\\
FRB\,20230222B & 187.8 & 27.7 & 0.11 & N & \cite{2025ApJS..280....6C}\\
FRB\,20230307A & 608.9 & 37.6 & 0.271 & N & \cite{2024Natur.635...61S}\\
FRB\,20230311A & 364.3 & 92.5 & 0.1918 & N & \cite{2025ApJS..280....6C}\\
FRB\,20230501A & 532.5 & 125.6 & 0.301 & N & \cite{2024Natur.635...61S}\\
FRB\,20230521B & 1342.9 & 138.8 & 1.354 & N & \cite{2025NatAs.tmp..131C}\\
FRB\,20230526A & $361.4\pm0.2$ & 50.0 & 0.157 & N & \cite{2024arXiv240802083S}\\
FRB\,20230626A & 451.2 & 39.2 & 0.327 & N & \cite{2024Natur.635...61S}\\
FRB\,20230628A & 345.15 & 39.1 & 0.1265 & N & \cite{2024Natur.635...61S}\\
FRB\,20230703A & 291.3 & 26.9 & 0.1184 & N & \cite{2025ApJS..280....6C}\\
FRB\,20230708A & $411.51\pm0.05$ & 50.0 & 0.105 & N & \cite{2024arXiv240802083S}\\
FRB\,20230712A & 586.96 & 39.2 & 0.4525 & N & \cite{2024Natur.635...61S}\\
FRB\,20230718A & $476.6\pm0.5$ & 393.0 & 0.0357 & N & \cite{2024ApJ...962L..13G}\\
FRB\,20230730A & 312.5 & 85.2 & 0.2115 & N & \cite{2025ApJS..280....6C}\\
FRB\,20230808F & 653.2 & 27 & 0.3472 & N & \cite{2025MNRAS.538.1800H}\\
FRB\,20230814A & 696.4 & 104.9 & 0.5535 & N & \cite{2025NatAs.tmp..131C}\\
FRB\,20230902A & $440.1\pm0.1$ & 34.0 & 0.3619 & N & \cite{2024arXiv240802083S}\\
FRB\,20230926A & 222.8 & 52.6 & 0.0553 & N & \cite{2025ApJS..280....6C}\\
FRB\,20231005A & 189.4 & 33.5 & 0.0713 & N & \cite{2025ApJS..280....6C}\\
FRB\,20231011A & 186.3 & 70.4 & 0.0783 & N & \cite{2025ApJS..280....6C}\\
FRB\,20231017A & 344.2 & 31.15 & 0.245 & N & \cite{2025ApJS..280....6C}\\
FRB\,20231025B & 368.7 & 48.6 & 0.3238 & N & \cite{2025ApJS..280....6C}\\
FRB\,20231120A & 438.9 & 43.8 & 0.07 & N & \cite{2024Natur.635...61S}\\
FRB\,20231123A & 302.1 & 89.7 & 0.0729 & N & \cite{2025ApJS..280....6C}\\
FRB\,20231123B & 396.7 & 40.2 & 0.2625 & N & \cite{2024Natur.635...61S}\\
FRB\,20231128A & 331.6 & 64.74 & 0.1079 & N & \cite{2025ApJS..280....6C}\\
FRB\,20231201A & 169.4 & 70 & 0.1119 & N & \cite{2025ApJS..280....6C}\\
FRB\,20231204A & 221 & 34.94 & 0.0644 & N & \cite{2025ApJS..280....6C}\\
FRB\,20231206A & 457.7 & 59.1 & 0.0659 & N & \cite{2025ApJS..280....6C}\\
FRB\,20231220A & 491.2 & 49.9 & 0.3355 & N & \cite{2025NatAs.tmp..131C}\\
FRB\,20231223C & 165.8 & 47.9 & 0.1059 & N & \cite{2025ApJS..280....6C}\\
FRB\,20231226A & $329.9\pm0.1$ & 145.0 & 0.1569 & N & \cite{2024arXiv240802083S}\\
FRB\,20231229A & 198.5 & 58.2 & 0.019 & N & \cite{2025ApJS..280....6C}\\
FRB\,20231230A & 131.4 & 23.03 & 0.0298 & N & \cite{2025ApJS..280....6C}\\
FRB\,20240114A & $527.7\pm0.1$ & 40.0 & 0.42 & Y & \cite{2024ApJ...977..177K}\\
FRB\,20240119A & 483.1 & 37.9 & 0.37 & N & \cite{2025NatAs.tmp..131C}\\
FRB\,20240123A & 1462 & 90.3 & 0.968 & N & \cite{2025NatAs.tmp..131C}\\
FRB\,20240201A & $374.5\pm0.2$ & 38.0 & 0.042729 & N & \cite{2024arXiv240802083S}\\
FRB\,20240209A & 176.57 & 55.5 & 0.1384 & Y & \cite{2025ApJ...979L..21S}\\
FRB\,20240210A & $283.73\pm0.05$ & 31.0 & 0.023686 & N & \cite{2024arXiv240802083S}\\
FRB\,20240213A & 357.4 & 40.1 & 0.1185 & N & \cite{2025NatAs.tmp..131C}\\
FRB\,20240215A & 549.5 & 48 & 0.21 & N & \cite{2025NatAs.tmp..131C}\\
FRB\,20240229A & 491.15 & 37.9 & 0.287 & N & \cite{2025NatAs.tmp..131C}\\
FRB\,20240310A & $601.8\pm0.2$ & 36.0 & 0.127 & N & \cite{2024arXiv240802083S}\\

\hline
\end{longtable}


\section{Modeling artificial neural network}\label{Appendix2}

\begin{figure}[htpb]
    \centering
    \includegraphics[scale=0.29]{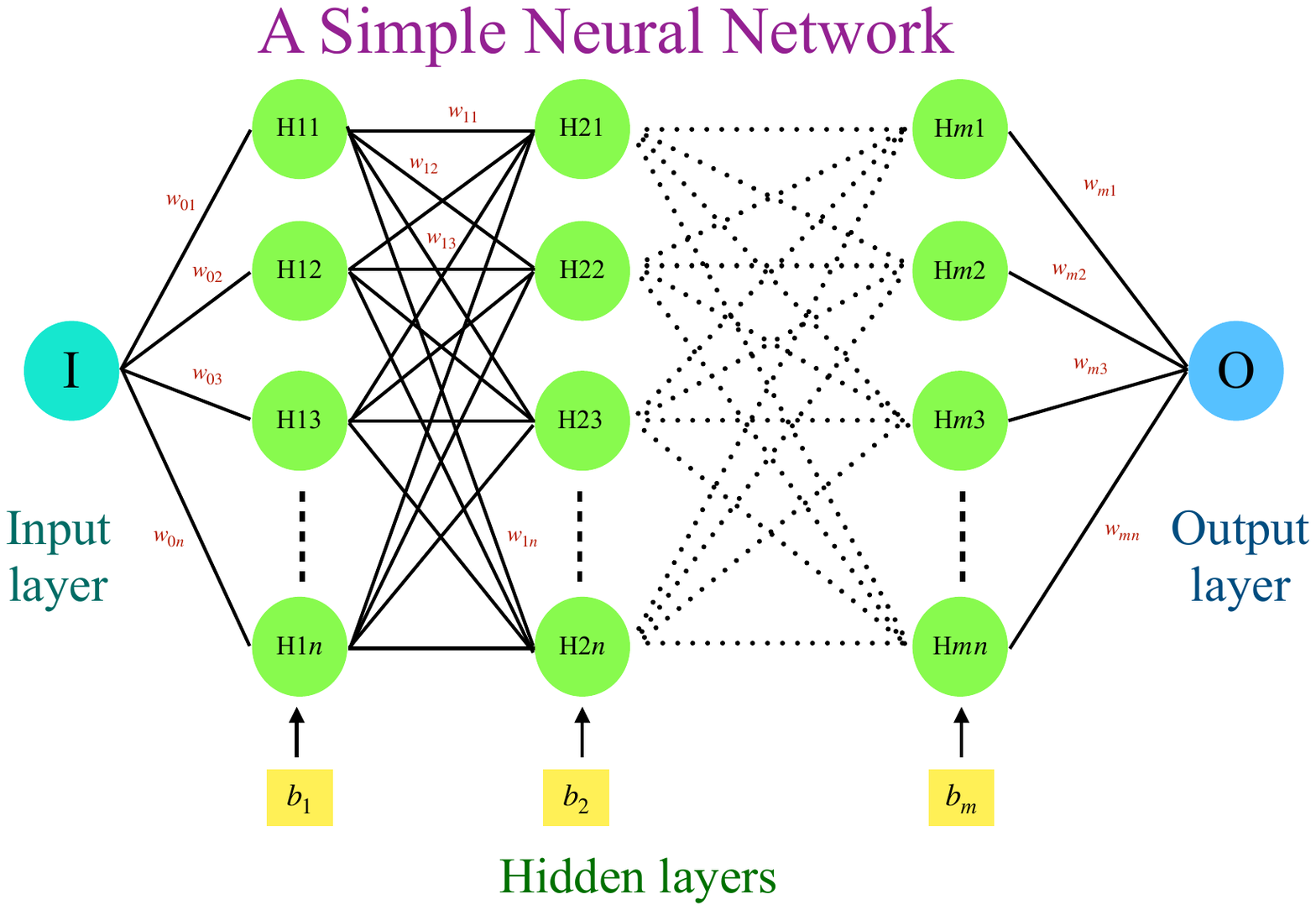}
    \caption{Schematic architecture of the feedforward neural network consisting of an input layer, $m$ fully connected hidden layers with $n$ neurons each, and a single output layer. Each hidden neuron performs a nonlinear transformation of the weighted sum of its inputs, enabling the network to learn complex, non-linear mappings.}
    \label{Fig: ANN}
\end{figure}

We implement a canonical feedforward ANN architecture to model the mapping between $z_\mathrm{s}$ and $\mathrm{DM}'$ as shown in Fig.~\ref{Fig: ANN}, where an input vector is first fed into one or more densely connected hidden layers. Our network architecture comprises an input layer, two fully connected hidden layers containing 100 neurons each, and a single output layer. Each neuron in the hidden layers computes a linear combination of its inputs, followed by the application of a nonlinear activation function, specifically ReLU. The transformed representations are propagated forward through successive layers, ultimately converging at the output node and yielding the final prediction of $\mathrm{DM}'$. 

The training process involves minimizing a weighted mean squared error (MSE) loss function, expressed as
\begin{align}
    \chi^2_\mathrm{MSE} = \frac{1}{\mathcal{N}}\sum_{i=1}^\mathcal{N}\frac{1}{\sigma_i^2}\left(\mathrm{DM}_\mathrm{pred} - \mathrm{DM}_{\mathrm{true},i}\right)^2,
\end{align}
where $\mathcal{N}$ is the number of training samples used in the neural network, $\mathrm{DM}_{\mathrm{true},i}$ is the true DM of the $i^\mathrm{th}$ input, and $\mathrm{DM}_\mathrm{pred}$ is its corresponding ML-predicted value. This formulation appropriately accounts for observational uncertainties by weighting $\sigma^2 = \sigma_{\mathrm{obs}}^2 + \sigma_{\mathrm{MW}}^2$, enhancing the robustness of the training procedure. Note that as $\sigma_{\mathrm{obs}} \ll \sigma_{\mathrm{MW}}$, we have $\sigma\approx\sigma_{\mathrm{MW}}$. Given the relatively limited size of the dataset, the data were partitioned using an 80:20 split, with 80\% allocated for training and 20\% for validation. This approach aligns with established ML protocols aimed at optimizing model generalization while maximizing the utilization of available samples.

Optimization of the network parameters, comprising weights and biases, is performed using the Adam optimizer, implemented within the \texttt{PyTorch} framework, with a fixed learning rate of 0.01, while prediction uncertainty is estimated by integrating Bayesian bootstrapping with Monte Carlo (MC) dropout. We choose the neural network architecture consisting of two hidden layers and incorporating dropout regularization with a probability of 0.09 (\texttt{torch.nn.Dropout}(0.09)). The network is trained multiple times on resampled datasets of $\mathrm{DM}'_i - z_{\mathrm{s},_i}$, where each data point was weighted by its measurement error, $\sigma_i$, to account for different observational uncertainties. Each bootstrapped ensemble captures a distribution of plausible models consistent with the data, while MC dropout introduces stochastic variability within each model to approximate Bayesian inference. By aggregating predictions from all networks, the method provides median values and confidence intervals that reflect both measurement noise and model uncertainty. This methodology provides a robust, data-driven approach for quantifying predictive uncertainty in neural network models, particularly suited for noisy astrophysical datasets where uncertainty quantification is as critical as predictive accuracy.

\begin{figure}[htpb]
    \centering
    \includegraphics[scale=0.53]{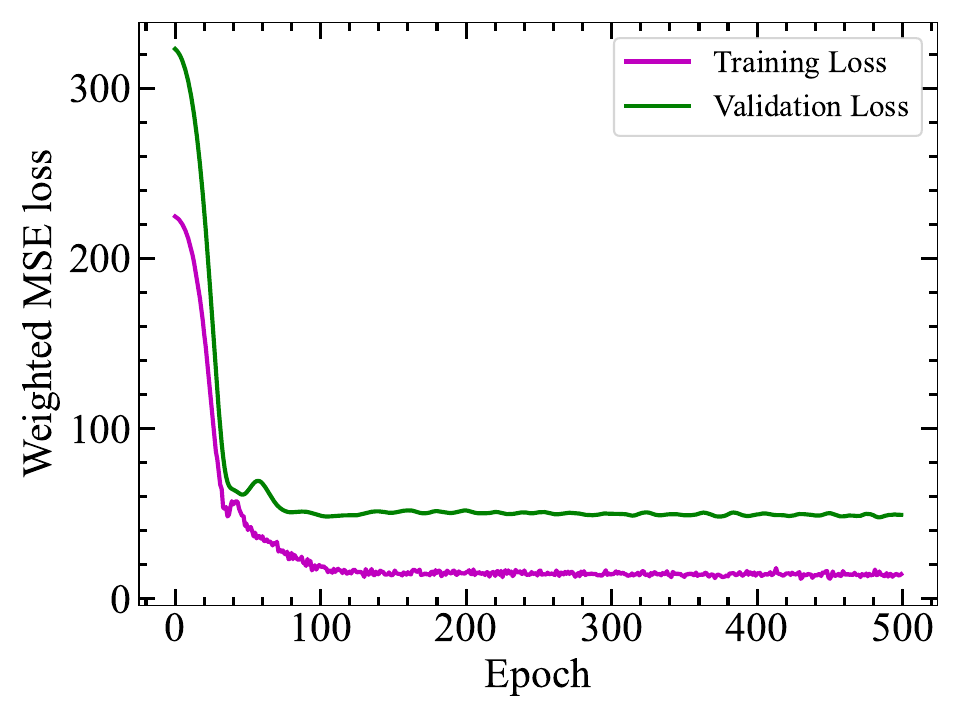}
    \caption{Training and validation loss as functions of epoch during the neural network optimization. The curves demonstrate the evolution of the MSE loss $\chi^2_\mathrm{MSE}$ with increasing number of epochs, indicating convergence and generalization performance.}
    \label{Fig: Loss_epoch}
\end{figure}

The progression of both training and validation losses, as depicted in Fig.~\ref{Fig: Loss_epoch}, indicates a monotonic decrease followed by saturation, demonstrating convergence and the model's capacity to generalize. The stability of our ML algorithm is evident from this figure, where the error decreases and saturates after sufficiently large epoch, indicating no overfitting. Following the completion of training, the neural network is employed to interpolate $\mathrm{DM}'$ values at intermediate redshifts that are not explicitly present in the training data. This interpolation enables an accurate estimation of $\mathrm{DM}'$ as a function of $z_\mathrm{s}$, consistent with the trends illustrated in Fig.~\ref{Fig: DM_z_ML}. Thus, the resulting model serves as a powerful, data-driven approach for inferring DM from redshift information in FRB studies.

\section{Probability densities of DM for MW halo, host, and IGM}\label{Appendix3}

Here we discuss the individual probability density functions that constitute the likelihood function in Eq.~\eqref{Eq: Likelihood3}. These distributions model DM of FRBs from the IGM, host galaxy, and halo. $\mathrm{DM}_\mathrm{IGM}$ is modeled by a non-Gaussian probability distribution motivated by both semi-analytic frameworks and hydrodynamical simulations, as described by \cite{2020Natur.581..391M}. Assuming $\Delta_\mathrm{IGM} = \mathrm{DM}_\mathrm{IGM}/\langle\mathrm{DM}_\mathrm{IGM}\rangle$, it can be defined as
\begin{align}
    P_\mathrm{IGM}\left(\Delta_\mathrm{IGM}\right) = A \Delta_\mathrm{IGM}^{-\beta_2} e^{-\frac{\left(\Delta_\mathrm{IGM}^{-\beta_1}-C_0\right)^2}{2\beta_1^2\sigma_\mathrm{DM}^2}},
\end{align}
where $\sigma_\mathrm{DM}$ is its standard deviation, $A$, $\beta_1$, $\beta_2$, and $C_0$ are free parameters characterizing the shape and width of the distribution. As noted in \cite{2020Natur.581..391M}, this form transitions smoothly to a Gaussian distribution in the regime of small $\sigma_\mathrm{DM}$ and has been validated against cosmological simulations for a wide range of halo masses and redshifts. For our analysis, we adopt these parameter values inferred by \cite{2021ApJ...906...49Z} based on the IllustrisTNG simulation.

The host galaxy DM contribution $\mathrm{DM}_\mathrm{Host}$ is inherently uncertain because of the complexity of the local galactic environments and limited observational constraints. Following \cite{2020Natur.581..391M}, we model this component using a log-normal distribution, given by
\begin{align}
    P_\mathrm{Host}\left(\mathrm{DM}_\mathrm{Host}\right) = \frac{1}{\sqrt{2\pi}\mathrm{DM}_\mathrm{Host}\sigma_\mathrm{Host}} e^{-\frac{\left(\ln{\mathrm{DM}_\mathrm{Host}}-\mu_\mathrm{Host}\right)^2}{2\sigma_\mathrm{Host}^2}},
\end{align}
where $e^{\mu_\mathrm{Host}}$ represents the median value with a variance $\left(e^{\sigma_\mathrm{Host}^2}-1\right)e^{2\mu_\mathrm{Host}+\sigma_\mathrm{Host}^2}$. This choice allows for a broad tail at high DM values, capturing scenarios where FRBs are embedded in dense environments such as HII regions or compact circumnuclear media. In line with previous analyses by \cite{2020Natur.581..391M}, we assume $e^{\mu_\mathrm{Host}} = 66\rm\,pc\,cm^{-3}$ and $\sigma_\mathrm{Host}=0.42\rm\,pc\,cm^{-3}$.

The DM distribution of halo $P_\mathrm{Halo}\left(\mathrm{DM}_{\mathrm{Halo}}\right)$ is modeled as a Gaussian distribution with mean $\mu_\mathrm{Halo}$ and standard deviation $\sigma_\mathrm{Halo}$, as follows
\begin{align}
     P_\mathrm{Halo}\left(\mathrm{DM}_{\mathrm{Halo}}\right) = \frac{1}{\sqrt{2\pi\sigma_\mathrm{Halo}^2}}e^{-\frac{\left({\mathrm{DM}_\mathrm{Halo}}-\mu_\mathrm{Halo}\right)^2}{2\sigma_\mathrm{Halo}^2}}.
\end{align}
Assuming $\mu_\mathrm{Halo} \approx 65\rm\,pc\,cm^{-3}$ and $\sigma_\mathrm{Halo}\approx5\rm\,pc\,cm^{-3}$, this parameterization closely reproduces the behavior of a uniform distribution over the interval $\mathcal{U}[50-80] \rm\,pc\,cm^{-3}$ \cite{2019MNRAS.485..648P} while offering a mathematically computable form suitable for likelihood computations.

\begin{figure}[htpb]
    \centering
    \subfigure[Case when $\Omega_\mathrm{b}$ is fixed. Here $H_{01}>H_{02}>H_{03}$.]{\includegraphics[scale=0.75]{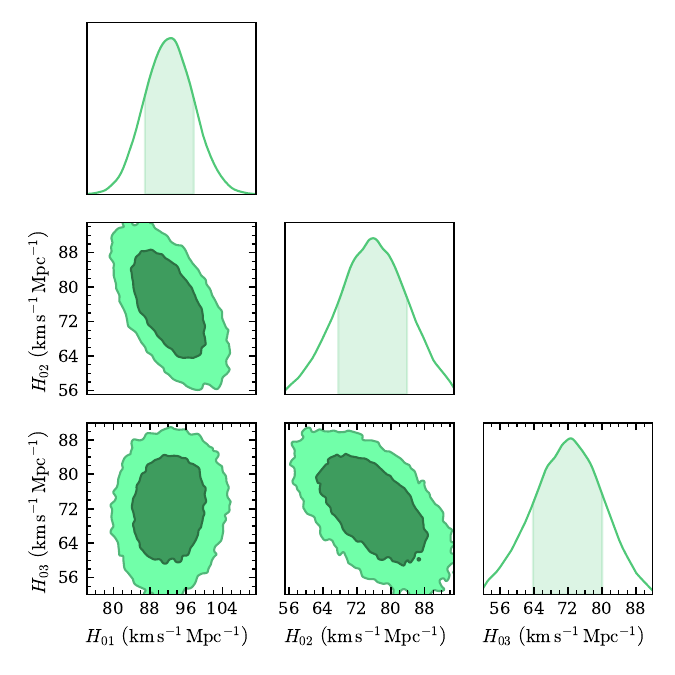}}    
    \subfigure[Case when $\Omega_\mathrm{b}h^2$ is fixed. Here $H_{01}<H_{02}<H_{03}$.]{\includegraphics[scale=0.75]{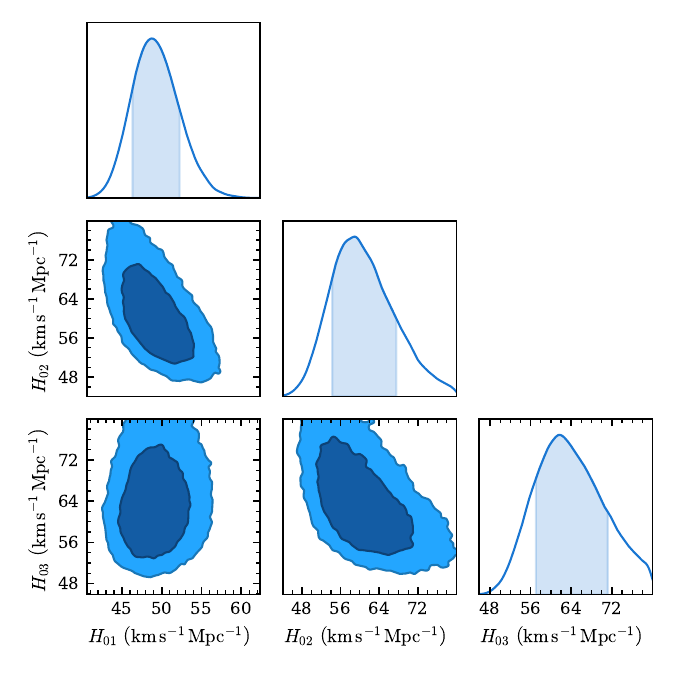}}    
    \caption{1$\sigma$ and 2$\sigma$ confidence regions of the joint likelihood of Equation~\eqref{Eq: Likelihood3} when it is optimized over Hubble constant values in different bins with $f_\mathrm{IGM}=0.85$. This plot is prepared with \texttt{ChainConsumer} function of Python.}
    \label{Fig: H0_omega}
\end{figure}

Together, these probability densities account for the dominant sources of uncertainty in modeling the total extragalactic DM. Owing to precise localization, the redshift measurements of the selected FRBs introduce negligible uncertainty, and are thus treated as fixed inputs in our statistical framework. Substituting these distribution functions into the likelihood function in Eq.~\eqref{Eq: Likelihood3}, we estimate the Hubble constant values in each bin using Eq.~\eqref{Eq: Macquart2} with the resulting distributions shown in Fig.~\ref{Fig: H0_omega}. Evidently, when $\Omega_\mathrm{b}$ remains fixed, $H_{01}>H_{02}>H_{03}$, whereas fixing $\Omega_\mathrm{b}h^2$ results in $H_{01}<H_{02}<H_{03}$.

\end{document}